\pdfoutput=1
\documentclass[aps,pra,reprint,superscriptaddress,preprintnumbers,nofootinbib,showpacs]{revtex4-1}

\usepackage{bm,graphicx,amsmath,amssymb,color,bbold,simplewick}
\usepackage[colorlinks=true, pdfstartview=FitV, linkcolor=blue, citecolor=blue, urlcolor=blue]{hyperref}

\graphicspath{{./1_Figures/}}

\DeclareMathOperator\ee{e}
\DeclareMathOperator\tr{tr}

\renewcommand{\Re}{\mathrm{Re}}
\newcommand{\der}{\partial}

\newcommand{\calO}{\mathcal{O}}
\newcommand{\dd}{\mathrm{d}}
\newcommand{\bep}{\begin{pmatrix}} 
\newcommand{\eep}{\end{pmatrix}}
\newcommand{\Sp}{\text{Sp}}
\newcommand{\SU}{\text{SU}}

\newcommand{\U}{\text{U}}
\newcommand{\1}{\mathbb{1}}
\newcommand{\RR}{\mathbb{R}}

\newcommand{\ZZ}{\mathbb{Z}}
\renewcommand{\epsilon}{\varepsilon}

\newcommand{\D}{\mathcal{D}}
\DeclareMathOperator*{\EE}{\mathbf{E}}
\newcommand{\rmt}{\text{RMT}}

\newcommand{\C}{\mathsf{C}}

\def\ba#1\ea{\begin{align}#1\end{align}}
\def\akakko#1{\left\langle #1 \right\rangle}
\def\mkakko#1{\left( #1 \right)}
\def\ckakko#1{\left\{ #1 \right\}}
\def\kkakko#1{\left[ #1 \right]}

\allowdisplaybreaks

\begin{document}

\preprint{RIKEN-QHP-203}

\title{Nonrelativistic Banks--Casher relation and random matrix theory for 
multi-component fermionic superfluids}

\author{Takuya~Kanazawa}
\affiliation{iTHES Research Group and Quantum Hadron Physics Laboratory,
RIKEN, Wako, Saitama 351-0198, Japan}
\author{Arata~Yamamoto}
\affiliation{Department of Physics, The University of Tokyo, Tokyo 113-0033, Japan}

\begin{abstract}
	We apply QCD-inspired techniques to study nonrelativistic $N$-component degenerate fermions 
	with attractive interactions.  By analyzing the singular-value spectrum 
	of the fermion matrix in the Lagrangian, we derive several exact relations 
	that characterize the spontaneous symmetry breaking $\U(1)\times\SU(N)\to\Sp(N)$ 
	through bifermion condensates. 
	These are nonrelativistic analogues of the Banks--Casher relation 
	and the Smilga--Stern relation in QCD. 
	Non-local order parameters are also introduced and 
	their spectral representations are derived, from which a nontrivial 
	constraint on the phase diagram is obtained. The effective theory of 
	soft collective excitations is derived and its equivalence to 
	random matrix theory  is demonstrated in the $\epsilon$-regime. 
	We numerically confirm the above analytical predictions in Monte Carlo simulations.
\end{abstract}

\maketitle

\section{\label{sc:intro}Introduction}

Spontaneous symmetry breaking is a universal concept across broad fields of physics.
The Bose--Einstein condensation of atoms is a marked example of quantum phenomena 
accessible in laboratory experiments \cite{Kapitza:1938,Osheroff:1972zz,Anderson:1995gf,Davis:1995pg}.
Superconductivity of electrons plays an essential role in condensed matter physics and furnishes diverse 
technological applications \cite{Bardeen:1957mv}.
Chiral symmetry breaking in quantum chromodynamics (QCD) is a dominant mechanism for mass generation in our universe \cite{Nambu:1960tm,Nambu:1961tp,Nambu:1961fr}.
The masses of elementary particles are generated by the Higgs mechanism \cite{Higgs:1964ia,Englert:1964et,Higgs:1964pj}.

Spontaneous symmetry breaking is driven by quantum effects.
For its exact derivation, the full information of a quantum 
many-body vacuum is necessary, 
but it is extremely difficult to obtain.
To tackle this difficult problem, many theoretical approaches have been developed in each field. 
Although they are formulated in different ways among different fields, 
the underlying physics must be common and 
an approach that proved successful in one field is expected to be applicable to another field.
Such an interdisciplinary endeavor is of vital importance 
to grasp the true nature of a universal phenomenon.

The target of this paper is spontaneous symmetry breaking in nonrelativistic multi-component degenerate fermions.  
This occurs in a variety of physical situations in nature.  
In nuclear physics, an atomic nucleus is composed of protons and neutrons with two spin states, 
entailing an approximate spin-isospin symmetry \cite{gaponov1996spin}. 
In ultracold atomic systems, $\SU(N)$-symmetric ultracold Fermi gases have been 
experimentally realized \cite{2010PhRvL.105s0401T}. 
The $\SU(N)$ Hubbard model on a lattice has also attracted 
attention \cite{2004PhRvL..92q0403H,2012NatPh...8..825T}.  
We refer to \cite{Wu:2003zzj,2004PhRvB..70i4521H,Veillette:2007zz,Cherng2007PRL,
Gorshkov2009,Cazalilla2009a,Szirmai2010a,Yip:2011zz,Hung2011a,Yip2013a,Zhang:2014xfa} 
for a partial list of works addressing the novel physics of 
multi-component Fermi gases, and \cite{Cazalilla:2014wfa} for a recent review. 

In this work, we apply analytical tools established in the study of spontaneous chiral symmetry breaking in QCD 
to interacting nonrelativistic fermions with an even number of components. 
As in QCD, we analyze the eigenvalues (more precisely, the singular values) 
of the fermion matrix in the Lagrangian formalism.%
\footnote{They should not be confused with the energy eigenvalues 
of the Hamiltonian operator in the Hamiltonian formalism.} 
The structure of the spectrum reflects realization of global symmetries in the ground state. 
We derive some \emph{exact} relations between the spectrum and symmetry breaking, 
including the nonrelativistic counterparts of the Banks--Casher relation (Sec.~\ref{sc:BC}) 
and the Smilga--Stern relation (Sec.~\ref{sc:SS}), both of which are well established in 
studies of the Dirac operator in QCD.  In addition, by relating two-point correlation functions 
of fermion bilinears to the singular-value spectrum, we show in Sec.~\ref{sc:u1} that if $\U(1)$ symmetry 
is spontaneously broken, then $\SU(N)$ symmetry must be broken down to $\Sp(N)$, and 
\emph{vice versa}, in $N$-component fermions.   
A salient feature of the Dirac spectrum in QCD is that it obeys random matrix theory (RMT) in a finite-volume regime 
called microscopic domain (or $\epsilon$-regime). In Sec.~\ref{sc:rmt} we derive the effective theory of 
soft collective excitations for nonrelativistic multi-component fermions, and identify the correspondence 
between the singular-value spectrum and RMT. 
We verify these analytical predictions by path-integral Monte Carlo simulations of nonrelativistic fermions 
on a lattice, utilizing powerful techniques developed in lattice QCD 
(Sec.~\ref{sc:NS}). 
In Appendixes, a few analytical derivations are given for completeness.

\section{\label{sc:BC}Banks--Casher-type relation}

Our main interest is in $N$-component degenerate fermions 
with $s$-wave contact interactions with $\U(N)$-symmetric theory, where $N=2,4,6,\dots$ is assumed to be \emph{even}. 
We will work in $D$-dimensional space with $D=2$ and $3$.  
The action in the imaginary-time formalism is given (in the unit $\hbar=1$) by
\ba
	S & = \int_x \kkakko{
	\sum_{i=1}^{N} \psi_i^* \mkakko{
	\der_\tau - \frac{\nabla^2}{2m}-\mu} \psi_i 
	+ \frac{c}{2}\,\bigg(\sum_{i=1}^{N}\psi^*_i\psi_i \bigg)^2 
	} 
	\label{eq:S}
\ea
with $\int_x\equiv  \int_0^\beta \dd \tau \int \dd^D \mathbf{x}$. 
The coupling $c<0$ ($c>0$) represents an attractive (repulsive) interaction, respectively.%
\footnote{In this paper, we ignore physics related to three-body interactions.}
The inverse temperature $\beta=1/k_BT$ is arbitrary at this stage. 
The partition function is given by the path integral  
$Z=\int \D \psi^*\D\psi\,\exp(-S)$.  
At $N=2$, Eq.~\eqref{eq:S} is reduced to the conventional 
spin-$1/2$ Fermi gas with $\U(1)\times\SU(2)$ symmetry. 

From here on, we concentrate on the attractive interaction and let $g\equiv -c>0$. 
By means of a Hubbard--Stratonovich transformation, one obtains 
$Z=\int \D \psi^*\D\psi \D\phi\,\exp(-S')$ with 
\ba
	S' & = \int_x \kkakko{ 
		\sum_{i=1}^{N} \psi_i^* \mkakko{
		\der_\tau - \frac{\nabla^2}{2m}-\mu-g\phi} \psi_i 
		+ \frac{g}{2}\phi^2
	}\, , 
	\label{eq:S'}
\ea
where $\phi(x)$ is a real bosonic auxiliary field. 
Now $S'$ is bilinear in fermion fields. 

If the system develops a fermion pair condensate $\akakko{\psi_i\psi_j}$, 
it breaks $\U(N)$ symmetry spontaneously. To extract the condensate, it is useful to add 
the following source term to the action 
\ba
	\delta S = - \frac{j}{2} \int_x \mkakko{
		\psi_i I_{ij} \psi_j + \text{h.c.}
	} \,,
	\label{eq:Jinsert}
\ea
with $I\equiv\bep 0 & 1 \\ -1 & 0 \eep\otimes \1_{N/2}$\,.  
This term breaks $\U(N)$ symmetry down to the unitary symplectic group defined by
\ba
	\Sp(N) & = \ckakko{  u\in\SU(N)\,|\,u^T I u = I  }\,. 
\ea 
We introduce the source term \eqref{eq:Jinsert} with $j>0$, and then let $j\to 0$ in the end of calculations.
A nonzero condensate in the $j\to 0$ limit signals spontaneous $\U(N)$ symmetry breaking.

Combining Eq.~\eqref{eq:Jinsert} with Eq.~\eqref{eq:S'} and going to 
the Nambu--Gor'kov representation, one finds
\ba
	& S'+\delta S 
	\notag
	\\
	=& \int_x \bigg[
	\sum_{k=1}^{N/2} \big(\begin{array}{cc}\psi^*_{2k-1} & \psi_{2k} \end{array}\big)
	\bep W & j \\ j & - W^\dagger \eep
	\bep\psi_{2k-1}\\\psi_{2k}^*\eep
	+ \frac{g}{2}\phi^2
	\bigg] 
\ea
with
\ba
	W \equiv \der_\tau - \frac{\nabla^2}{2m}-\mu-g\phi \,.
	\label{eq:W}
\ea
The next step is to integrate out fermions, with the result 
\ba
	Z(j) & = \int \D \phi ~{\det}^{N/2} \bep W & j \\ j & - W^\dagger \eep 
	\exp\mkakko{-\frac{g}{2}\int_x~\phi^2}
	\notag
	\\
	& = \int \D \phi ~{\det}^{N/2}\mkakko{j^2+WW^\dagger}
	\exp\mkakko{-\frac{g}{2}\int_x~\phi^2}\,. 
	\label{eq:Zf}
\ea
This form manifestly shows that the path-integral measure is 
positive definite so that this theory can be 
simulated with standard Monte Carlo methods. 
We warn that this is no longer true if $N$ is odd or if the interaction is repulsive. 

It is now straightforward to find the fermion condensate by taking the derivative with $j$,
\ba
	\frac{1}{2}\akakko{\psi^T I \psi+\text{h.c.}} 
	& = \lim_{V\to\infty}\frac{1}{\beta V} 
	\frac{\dd}{\dd j} \log Z(j)
	\notag
	\\
	& = \frac{N}{2} \lim_{V\to\infty}\frac{1}{\beta V} 
	\bigg\langle 
		\sum_{n}\frac{2j}{j^2+\Lambda_n^2}
	\bigg\rangle 
	\notag
	\\
	& = \frac{N}{2} \int_0^\infty \!\!\! \dd\Lambda~\frac{2j}{j^2+\Lambda^2} R_1(\Lambda),
	\label{eq:R11}
\ea
where $V$ is the spatial volume and $\Lambda_n \geq 0$ are square roots of the eigenvalues of 
$WW^\dagger$ (i.e., the \emph{singular values} of $W$).  
The \emph{spectral density} (or one-point function), $R_1(\Lambda)$, is defined 
for $\Lambda \geq 0$ as
\ba
	R_1(\Lambda) & \equiv \lim_{V\to\infty}\frac{1}{\beta V} 
	\bigg\langle 
		\sum_{n}\delta(\Lambda-\Lambda_n)
	\bigg\rangle 
	\label{eq:R1}
\ea
where the average $\akakko{\cdots }$ is taken with respect to the measure \eqref{eq:Zf}. 
By taking the limit $j\to 0$, we arrive at 
\ba
	 \lim_{j\to +0} \frac{1}{2}\akakko{\psi^T I \psi+\text{h.c.}} 
	  = \frac{N}{2} \pi \lim_{j\to +0} R_1(0) .
	\label{eq:newBC}
\ea
This relation, linking the density of small singular values of $W$  
to spontaneous symmetry breaking $\U(1)\times\SU(N)\to\Sp(N)$,%
\footnote{For $N=2$, the breaking pattern is $\U(1)\to \varnothing$ since $\Sp(2)\cong \SU(2)$.} 
is the main result of this section.   
This is a generalization of the celebrated Banks--Casher relation  
for gauge theories \cite{Banks:1979yr} to nonrelativistic fermions. 
Several remarks are in order.  
\begin{itemize}
	\item 
	As is clear from the derivation above, 
	the new relation \eqref{eq:newBC} holds both in the normal 
	phase and in the superfluid phase. The temperature, chemical potential 
	and the interaction strength are arbitrary. 
	\item 
	The action \eqref{eq:S} based on the $s$-wave 
	contact interaction has an intrinsic short-distance 
	cutoff scale (i.e., the effective range of the inter-particle potential). 
	This implies that it is not physically meaningful to integrate over 
	$\Lambda$ up to infinity 
	in Eq.~\eqref{eq:R11} beyond the short-distance cutoff. 
	However, a more elaborate treatment of the integral would not 
	change the final formula \eqref{eq:newBC} because   
	all contributions to Eq.~\eqref{eq:R11} from regions 
	away from the origin will eventually drop out in the limit $j\to 0$. 
	Thus Eq.~\eqref{eq:newBC} holds irrespective of the detailed  
	short-distance physics. 
	\item 
	We stress that the positivity of the measure is essential 
	for the derivation of Eq.~\eqref{eq:newBC}. 
	If the measure becomes negative or complex, the 
	spectral density tends to be a violently oscillating function that has 
	no smooth thermodynamic limit 
	\cite{Akemann:2004dr,Osborn:2005ss,Kanazawa:2011tt,
	Kanazawa:2014lga,Verbaarschot:2014upa}, so that 
	the last step from Eq.~\eqref{eq:R11} to Eq.~\eqref{eq:newBC} replacing 
	$\frac{2j}{j^2+\Lambda^2}$ with $2\pi \delta(\Lambda)$ 
	is invalidated. This suggests that this kind of an exact formula 
	will not exist in a spin-imbalanced Fermi gas, 
	even though the condensate itself may exist.   
	\item 
	In the free limit $g\to 0$, one can compute $R_1(\Lambda)$ analytically, as outlined in \autoref{ap:free}. 
	In $D=3$ dimensions at $T=0$, we find
	\ba
		R_1(\Lambda)\propto \Lambda^{3/2}
	\ea
	for $\mu=0$ and  
	\ba
		R_1(\Lambda)\propto \sqrt{\mu}\,\Lambda 
		\label{eq:Rlinear}
	\ea
	for $\mu\gg \Lambda>0$. In either case $R_1(0)=0$ gives a 
	vanishing condensate, but it is worthwhile to note 
	that the density of small eigenvalues is substantially 
	enhanced for $\mu>0$ as compared to $\mu=0$. 
	This means that a positive chemical potential 
	(or the presence of a Fermi surface) acts as a catalyst of 
	spontaneous symmetry breaking. Analogous phenomena 
	occur in the singular-value spectrum 
	of the Dirac operator in dense QCD-like theories 
	\cite{Kanazawa:2011tt} 
	and the Dirac spectrum of QCD in an external magnetic field 
	\cite{Shushpanov:1997sf}; in both cases the spectral density  
	near the origin is enhanced from $\sim \Lambda^3$ to $\sim \Lambda$. 
	\item
	While the above derivation focuses on the 
	$\Sp(N)$-symmetric condensate $\akakko{\psi^T I \psi}$, 
	one can also consider   
	$\akakko{\psi^T I T^A \psi}$ and $\akakko{\psi^\dagger t^a \psi}$, 
	where \cite{Peskin:1980gc,Kogut:2000ek}
	\begin{itemize}
		\item[$\diamond$] 
		$\{T^A\}\cdots$$N(N-1)/2-1$ generators of the coset space 
		$\SU(N)/\Sp(N)$, normalized as $\tr(T^A T^B)=\frac{1}{2}\delta^{AB}$.  
		$(T^A)^TI=IT^A$ holds. 
		\item[$\diamond$] 
		$\{t^a\}\cdots$$N(N+1)/2$ generators of $\Sp(N)$, normalized as 
		$\tr(t^a t^b)=\frac{1}{2}\delta^{ab}$. $(t^a)^TI=-It^a$ holds. 
	\end{itemize}
The former condensate transforms in the rank-2 antisymmetric tensor 
representation of $\Sp(N)$, while the latter in the adjoint representation of $\Sp(N)$. 
From the Vafa-Witten theorem \cite{Vafa:1983tf,Kosower:1984aw},  one can show $\akakko{\psi^T I T^A \psi}=\akakko{\psi^\dagger t^a \psi}=0$ for any $j\ne 0$. 
This argument assures that $\Sp(N)$ symmetry is unbroken for any $j\ne 0$. Namely, $\Sp(N)$-symmetric states have lower free energy than $\Sp(N)$-breaking states at $j\ne 0$. Then, if any 
$\Sp(N)$-breaking states are degenerate with 
$\Sp(N)$-symmetric states in the $j\to 0$ limit, 
$\Sp(N)$ symmetry could be spontaneously broken.
We will assume that $\Sp(N)$ is not spontaneously broken throughout the remainder of this paper. 
\end{itemize}

\section{\label{sc:u1}\boldmath $\U(1)$ versus $\SU(N)$ symmetry}
	
\begin{figure}[h]
		\includegraphics[width=.4\textwidth]{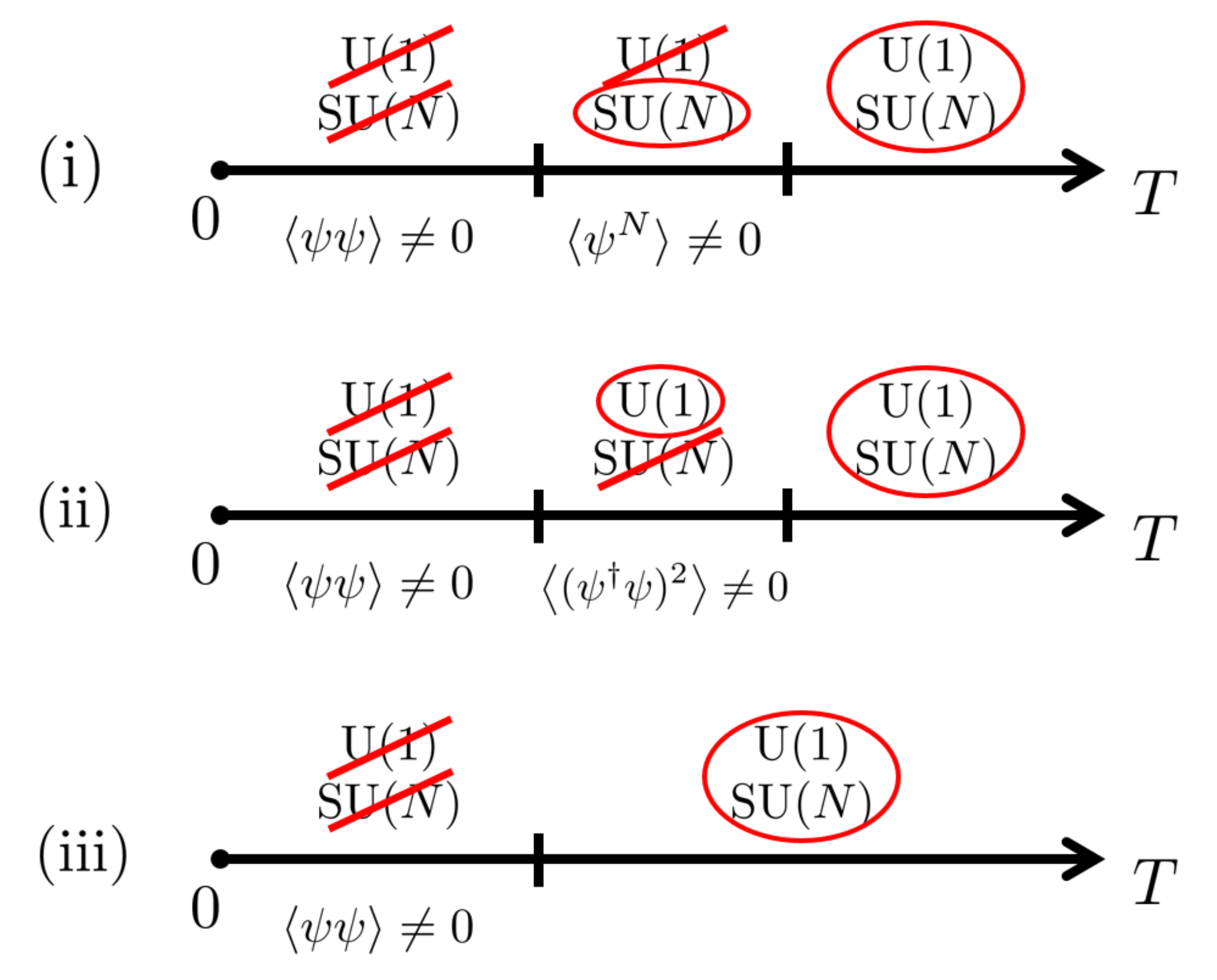}
		\caption{\label{fg:3phases} 
		Classification of possible finite-temperature phase diagrams 
		for even $N\geq 4$. }
	\end{figure}

	While $\akakko{\psi^T I \psi}\ne 0$ signals spontaneous breakdown 
	of \emph{both} $\U(1)$ and $\SU(N)$ for even $N\geq 4$, 
	one can in principle 
	also imagine a phase where either $\U(1)$ or $\SU(N)$ is broken 
	but the other is unbroken. 
	Taking such intermediate phases into account leads us to three 
	distinct phase diagrams sketched in Fig.~\ref{fg:3phases}. 
	In cases (i) and (ii) there appear phases with partial symmetry breaking, while 
	in case (iii) $\U(1)$ and $\SU(N)$ are simultaneously restored. 
	(Similar diagrams can be drawn for a varying interaction strength.) 	
	
	In this section, we shall use spectral methods inspired by QCD 
	to argue that 
	such exotic intermediate phases should not arise at least for $N=4$. 
	The key requirement in our analysis is that, to characterize phases with no bilinear condensate, one 
	must consider higher-order condensates containing more than two fermions, 
	as a source of symmetry breaking. We clarify the necessary 
	and sufficient condition for the singular-value spectrum of $W$ to support 
	such higher-order condensates in a phase with $\lim_{j\to 0} R_1(0)=0$. 
	
	We mention that there are ample literature on symmetry breaking driven by 
	higher-order condensates in high-energy physics. In QCD at finite density, 
	the breaking of $\U(1)$ baryon number symmetry and chiral symmetry in 
	color-superconducting phases is characterized by a six-quark condensate 
	and a four-quark condensate, respectively 	\cite{Rajagopal:2000wf,Alford:2007xm}. 
	Four-quark condensates also appear in the hypothetical Stern phase of QCD 
	\cite{Stern:1998dy,Kogan:1998zc,Kanazawa:2015kca}. 
	Non-local four-quark operators play a central role in the debate over 
	effective restoration of the anomalous $\U(1)_A$ symmetry 
	at high temperature \cite{Shuryak:1993ee,Cohen:1996ng,
	Bernard:1996iz,Chandrasekharan:1998yx,Edwards:1999zm,Kanazawa:2015xna}. 
	Furthermore, in some inhomogeneous phases 
	of QCD, the bilinear condensate is washed out by strong fluctuations of 
	phonons, so the leading condensate consists of four quarks \cite{Hidaka:2015xza}  
	(see \cite{radzihovsky2011a,radzihovsky2011b} for analogs in condensed matter physics). 
	
\begin{figure}[h]
		\includegraphics[width=.25\textwidth]{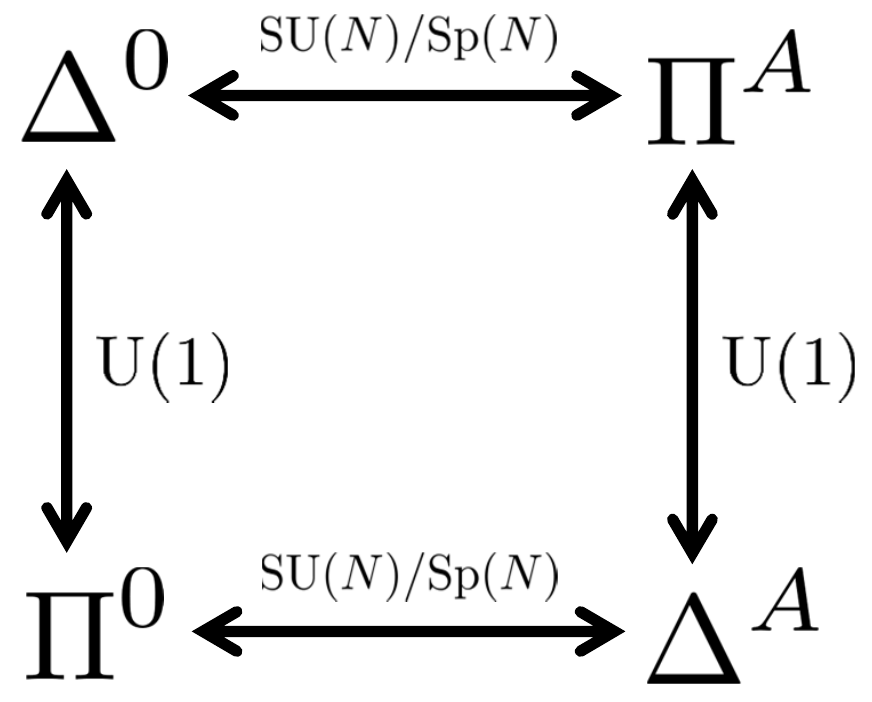}
		\caption{\label{fg:PiDelta} 
		The four fermion bilinears that transform to each other under $\U(1)$ 
		and $\SU(N)/\Sp(N)$ rotations.}
\end{figure}

Returning to the nonrelativistic $N$-component system of fermions, 
we define four bilinears as
\begin{subequations}
	\ba
		\Pi^0(x) & \equiv i( \psi^T I T^0 \psi + \psi^\dagger T^0 I \psi^* ) 
		\\
		\Delta^0(x) & \equiv \psi^T I T^0 \psi - \psi^\dagger T^0 I \psi^* 
		\\
		\Pi^A(x) & \equiv i( \psi^T I T^A \psi + \psi^\dagger T^A I \psi^* ) 
		\\
		\Delta^A(x) & \equiv \psi^T I T^A \psi - \psi^\dagger T^A I \psi^*  
	\ea
	\label{eq:multiplet}%
\end{subequations}
where $\{T^A\}$ are the generators of $\SU(N)/\Sp(N)$ as before, 
and $T^0\equiv \1_N/\sqrt{2N}$. These operators 
are mixed with each other 
under $\U(1)\times \SU(N)$ transformations, as summarized in Fig.~\ref{fg:PiDelta}. 
We define the integrated connected correlator of a field $X=\{\Pi^0,\Delta^0,\Pi^A,\Delta^A\}$ as 
\ba
	\C_X \equiv \int_x\int_y 
	\big\{
		\akakko{X(x)X(y)}-\akakko{X(x)}\akakko{X(y)}
	\big\} \,,
	\label{eq:cx}
\ea
where the averages are taken with respect to the measure \eqref{eq:Zf}.
This is an extensive quantity and must be divided by $\beta V$ when the thermodynamic limit is taken later.
The explicit forms of $\C_X$ are presented in \autoref{ap:cor}.

	Let us introduce non-local observables that are sensitive to 
	the realization of $\U(1)$ and $\SU(N)$ symmetry. 
	Since $\Pi^0$ mixes with $\Delta^A$ under 
	$\SU(N)$ transformations [cf.~Fig.~\ref{fg:PiDelta}], 
	one must have
	\ba
		\C_{\Pi^0} = \C_{\Delta^A}  
	\ea
	in the $j\to 0$ limit if $\SU(N)$ is unbroken.  
	This property prompts us to define 
	\ba
	\begin{split}
		\omega_{\scriptscriptstyle\SU(N)} & \equiv 
		\frac{1}{\beta V} \sum_A (\C_{\Pi^0} - \C_{\Delta^A})
		\\
		& = \frac{1}{\beta V} \kkakko{
			\frac{N^2-N-2}{2}\C_{\Pi^0} - \sum_A \C_{\Delta^A}
		}
		\\
		& = \frac{2(N^2-N-2)}{\beta V}
		\bigg\langle \!\tr \frac{j^2}{(W^\dagger W+j^2)^2} \bigg\rangle
		\\
		& = 2(N^2-N-2) 
		\int_0^\infty\!\!\! \dd\Lambda \frac{j^2}{(\Lambda^2+j^2)^2} R_1(\Lambda)\,,
		\label{eq:omegaSUN}
	\end{split}
	\ea
	where formulas in \autoref{ap:cor} have been used repeatedly.
	Next, Fig.~\ref{fg:PiDelta} shows that $\Pi^A$ and $\Delta^A$ mix 
	with each other under $\U(1)$ transformations. Hence 
	one must have
	\ba
		\C_{\Pi^A} = \C_{\Delta^A}
	\ea
	in a phase with unbroken $\U(1)$ symmetry. Let us define 
	\ba
	\begin{split}
		\tilde\omega_{\scriptscriptstyle\U(1)} & \equiv 
		\frac{1}{\beta V}\sum_A 
		( \C_{\Pi^A} - \C_{\Delta^A} )
		\\
		& = 2(N^2-N-2) 
		\int_0^\infty\!\!\! \dd\Lambda \frac{j^2}{(\Lambda^2+j^2)^2} R_1(\Lambda)\,. 	 
	\end{split}
	\ea
	Intriguingly, this is exactly equal to Eq.~\eqref{eq:omegaSUN}. Hence  
	\ba
		\omega_{\scriptscriptstyle\SU(N)} 
		= 
		\tilde\omega_{\scriptscriptstyle\U(1)}
		\label{eq:omom}
	\ea  
	follows.
        What is the physical meaning of this relation?
	Let us consider the following two cases separately.
\begin{itemize}
	\item $N\geq 6$. 
	Since 
	\ba
		\qquad 
		\tilde\omega_{\scriptscriptstyle\U(1)} \! &  = 
		- \frac{2}{\beta V} \! 
		\int_{x,y} \!\!\!\!   
		\akakko{\psi^T(x) I T^A \psi(x)\psi^T(y) I T^A \psi(y)} 
		\notag
		\\
		& \quad + \text{h.c.} 
		\notag
	\ea
	is a charge-4 condensate, it must vanish when 
	$\ZZ_N\subset\SU(N)$ with $N\geq 6$  
	is restored, irrespective of the $\U(1)$ symmetry realization.  
	In other words, unbroken $\SU(N)$ is enough to ensure the degeneracy of 
	$(\Pi^0,\Delta^0,\Pi^A,\Delta^A)$ even though $\U(1)$ could still 
	be broken by higher order condensates. 
	Thus Eq.~\eqref{eq:omom} does \emph{not} tell us anything about the interrelation 
	between $\U(1)$ and $\SU(N)$ symmetries --- we only learn that 
	the restoration of $\SU(N)$ symmetry requires not only $\lim_{j\to 0} R_1(0)=0$ but also 
	\ba
		\lim_{j\to 0}\int_0^\infty\!\!\! \dd\Lambda 
		\frac{j^2}{(\Lambda^2+j^2)^2} R_1(\Lambda)=0\,,
	\ea
	which is a far more stringent condition than $\lim_{j\to 0} R_1(0) =0$.%
	\footnote{If $\lim_{j\to 0} R_1(0) >0$, 
	then $\omega_{\scriptscriptstyle\SU(N)}$ and 
	$\tilde\omega_{\scriptscriptstyle\U(1)}$ blow up to infinity 
	as $j\to 0$. This is attributed to the IR divergence caused by 
	the coupling of $\Pi^0$ and $\Pi^A$ to the gapless Nambu--Goldstone modes.} 
	\item $N=4$. 
	Unbroken $\SU(4)$ symmetry 
	does not imply $\tilde\omega_{\scriptscriptstyle\U(1)}=0$, so  
	$\tilde\omega_{\scriptscriptstyle\U(1)}$ can now be treated 
	as a faithful order parameter for $\U(1)$ symmetry breaking. 
	We interpret the coincidence \eqref{eq:omom} as an indication that 
	$\U(1)$ breaking goes hand-in-hand with 
	$\SU(4)$ breaking. 
	Hence intermediate phases as depicted in Fig.~\ref{fg:3phases} 
	are not expected to arise in the phase diagram.  
\end{itemize}
	Since there is no obvious reason to regard the $N=4$ fermion system 
	as exceptional, we conjecture that the simultaneous restoration of 
	$\U(1)$ and $\SU(N)$ would be a generic phenomenon for $N\geq 4$. 
	A further investigation on this issue is left for future work.   
	
	Finally we wish to analyze the possibility that 
	\emph{both} $\U(1)$ and $\SU(N)$ are broken by higher-order condensates despite $\akakko{\psi^T I \psi}=0$. 
	This hypothetical phase, characterized by $\lim_{j\to 0} R_1(0)=0$ 
	\emph{and} 
	\ba
		\lim_{j\to 0}\int_0^\infty\!\!\! \dd\Lambda 
		\frac{j^2}{(\Lambda^2+j^2)^2} R_1(\Lambda) >0\,, 
		\label{eq:Rco}
	\ea
	is not excluded by the arguments in this section.%
	\footnote{This kind of exotic symmetry breaking seems to occur 
	in the Stern phase of QCD \cite{Stern:1998dy} 
	and the Fulde--Ferrell--Larkin--Ovchinnikov phase of 
	imbalanced fermions, 
	where the bilinear condensate is unstable and superfluidity is driven by a quartic 
	condensate \cite{radzihovsky2011a,radzihovsky2011b}. It must be warned,  
	however, that the path-integral measure of imbalanced fermions is 
	not positive definite and $R_1(\Lambda)$ will not be a smooth positive 
	function of $\Lambda$.}  
	What is the form of $R_1(\Lambda)$ consistent with Eq.~\eqref{eq:Rco}?  
	It is readily seen that if $R_1(\Lambda)$ is strictly zero in the range $0\leq \Lambda\leq \lambda_0$ 
	for some $\lambda_0>0$ (as is the case for free fermions 
	at finite temperature), then 
	$\omega_{\scriptscriptstyle\SU(N)}=\tilde\omega_{\scriptscriptstyle\U(1)}
	=\calO(j^2)\to 0$ and the symmetry is restored. 
	Thus a nonzero density of eigenvalues 
	in the infinitesimal vicinity of the origin is a necessary condition for 
	Eq.~\eqref{eq:Rco}.  More precisely, Eq.~\eqref{eq:Rco} holds  
	if $R_1$ has the form%
	\footnote{$R_1(\Lambda)\sim j^2\delta(\Lambda)$ yields 
	$\omega\ne 0$, too, but such a singular form does not seem to be  
	physically well motivated.}
	\ba
		R_1(\Lambda) & \sim j^{\alpha} \Lambda^{1-\alpha}
		\qquad \text{for} \quad 0\leq\alpha\leq 1\,.
		\label{eq:Rexa}
	\ea
	A somewhat puzzling instance of the behavior \eqref{eq:Rexa} 
	is encountered in a free theory at $T=0$, where 
	$R_1(\Lambda)\propto \Lambda$ for $\mu>0$ (see \autoref{ap:free}). 
	Our interpretation is that this is not a true 
	symmetry breaking but rather an indication that free fermions 
	at $\mu>0$ is on the verge of symmetry breaking.
	At $\mu>0$, a nonzero density of states at the Fermi surface 
	ensures that fermion  pairs condense and 
	break symmetries spontaneously for an arbitrarily weak 
	attractive interaction $g>0$, i.e., the Cooper instability. We believe that 
	$\omega_{\scriptscriptstyle\SU(N)}=\tilde\omega_{\scriptscriptstyle\U(1)}\ne 0$ 
	at $g=0$ should be seen as an extrapolation of symmetry breaking in the limit $g\to +0$.  
	Note that they vanish as soon as we raise the 
	temperature from zero; namely, the true many-body effect is 
	needed to achieve $\omega_{\scriptscriptstyle\SU(N)}=\tilde\omega_{\scriptscriptstyle\U(1)}\ne 0$ 
	at any small but nonzero $T>0$. A quite similar phenomenon is known to occur when Dirac 
	fermions are subjected to an external magnetic field 
	in $2+1$ dimensions: the chiral condensate assumes a nonzero 
	value even in a free theory \cite{Gusynin:1994re,Gusynin:1994va}.    
	This deceiving condensate evaporates at any nonzero 
	temperature \cite{Das:1995bn}, similarly to our case.

\section{\label{sc:SS}Smilga--Stern-type relation}

One of the defining features of superfluidity 
is a nonzero stiffness (helicity modulus) \cite{Fisher:1973zzc}.%
\footnote{The helicity modulus is nothing but the squared pion decay constant 
in the terminology of QCD literature \cite{Hasenfratz:1989pk}.} 
It is important to understand how the information of the stiffness 
is imprinted in the spectral density $R_1(\Lambda)$. 
In this section we apply 
the method of low-energy effective field theory (EFT) to show that, 
while $\lim_{j\to 0}R_1(0)$ is proportional to the condensate, the \emph{slope} 
of $\lim_{j\to 0}R_1(\Lambda)$ 
is sensitive to the phase stiffness. This is a generalization of the so-called 
Smilga--Stern relation \cite{Smilga:1993in,Osborn:1998qb,Toublan:1999hi} 
in QCD to nonrelativistic superfluids. 
Our method is applicable to even $N\geq 4$ 
in the phase where $\akakko{\psi^T I \psi}\ne0$. This requires 
$D=3$ at sufficiently low $T$ or $D=2$ at $T=0$.  

EFT is a powerful method enabling a systematic description of 
low-energy physics based on symmetries. It can be equally 
applied to systems with or without Lorentz invariance, as has been theoretically 
demonstrated in \cite{Leutwyler:1993gf,Watanabe:2012hr,Hidaka:2012ym,Takahashi:2014vua};   
see \cite{Watanabe:2014fva,Andersen:2014ywa} for a comprehensive overview of the subject. 
In multi-component Fermi gases with even $N\geq 4$, fermions are gapped through $s$-wave pairing 
and the dominant excitations at low energy are gapless Nambu--Goldstone modes 
originating from the symmetry breaking $\U(1)\times \SU(N)\to \Sp(N)$. 
Since the construction of the effective Lagrangian in this case closely parallels 
previous works in two-color QCD \cite{Peskin:1980gc,Smilga:1994tb,Kogut:1999iv,Kogut:2000ek,
Kanazawa:2009ks,Kanazawa:2011tt}, we refer to these references 
for details and only recapitulate the main ideas. 

The first step is to generalize the source term \eqref{eq:Jinsert} to 
\ba
	\delta S = - \frac{1}{2} \int_x (
		\psi^T J \psi + \text{h.c.}
	)
	\label{eq:SJgeneral}
\ea
where
\ba
		J & = jI + \sum_{A} j_A IT^A 
		\label{eq:defJ}
\ea
is the most general decomposition of an antisymmetric $N\times N$ 
matrix \cite{Kanazawa:2011tt}.
Corrections to the effective action 
due to $J$ can be sorted out in a perturbative manner. At leading 
order in the number of derivatives 
($\der_\tau$, $\nabla$) and the external field ($J$) we obtain 
\ba
	\begin{split}
		\mathcal{L}_{\rm eff} & = 
		F^2\!\kkakko{
			\tr\!\mkakko{\der_\tau \Sigma^\dagger\der_\tau \Sigma}
			+ v^2 \tr\!\mkakko{\nabla_i \Sigma^\dagger \nabla_i \Sigma} 
		}
		\\
		& \quad + \frac{1}{2}\!\kkakko{
			(\der_\tau\phi)^2 + \tilde{v}^2(\nabla_i \phi)^2 
		}
		+ \Phi\,\Re\tr(J \tilde\Sigma)\,, 
	\end{split}
	\label{eq:Lmain}
\ea 
which will be valid if $k_{\rm B}T$ is much lower than the gap in the 
single-particle excitation spectrum.  

Several remarks are in order. 
\begin{itemize}
	\item 
	The coset manifold $\SU(N)/\Sp(N)$ is parametrized by  
	$\Sigma(x) = UIU^T=U^2I$ \cite{Kogut:2000ek} with 
	\ba
		U(x) & =\exp\mkakko{i\frac{\pi^A(x)T^A}{2F}}, 
	\ea
	where $\{\pi^A\}$ are the Nambu--Goldstone modes. 
	$\Sigma$ satisfies $\Sigma^T=-\Sigma$ and $\Sigma^\dagger \Sigma=\1_N$. 
	The coefficient of $(\der_\tau \pi^A)^2$ in Eq.~\eqref{eq:Lmain} is normalized to $1/2$. 
	$v$ denotes the velocity of the $\pi$ fields. 
	\item 
	The superfluid phonon is represented by $\phi(x)$, with the velocity $\tilde{v}$.
	In $\tilde\Sigma$, the phonon is coupled to $\pi$ as
	\ba
		\tilde\Sigma & = \Sigma \ee^{i\phi/f} \,. 
	\ea
        In two-color QCD, $\phi(x)$ is absent because 
	the axial $\U(1)$ symmetry in QCD is violated by chiral anomaly.
	\item 
	The last term in Eq.~\eqref{eq:Lmain} containing $J$ breaks the $\SU(N)$ symmetry 
	explicitly and generates a nonzero gap (``mass'') for the Nambu--Goldstone modes. 
	At $^\forall j_A=0$ we have  
	\ba
		m_\pi^2 =\frac{j\Phi}{2F^2}  \quad \text{and} \quad  
		m_{\phi}^2 = \frac{jN\Phi}{f^2}\,. 
		\label{eq:NGmass}
	\ea
	\item 
	Evaluating the derivative of $\log Z$ with $j$ at $^\forall j_A=0$ 
	one finds
	\ba
		\Phi & = \lim_{j\to 0} \frac{1}{2N} \akakko{\psi^T I \psi+\text{h.c.}} \,. 
	\ea
	Combined with our Banks--Casher-type relation \eqref{eq:newBC}, 
	this means $\lim_{j\to 0} R_1(0)=2\Phi/\pi$. 
	We note that $F, f, v, \tilde{v}$ and $\Phi$ all depend implicitly on $T, \mu, g$ and $N$.   
	\item 
	Generally, in the absence of Lorentz invariance, terms linear in the time derivative can appear 
	in effective Lagrangians and modify dispersion relations of Nambu--Goldstone modes 
	qualitatively \cite{Leutwyler:1993gf,Watanabe:2012hr,Hidaka:2012ym}.  
	This indeed occurs in the three-component fermionic superfluids \cite{He:2006ne,Yip:2011zz}. 
	However this does not occur for even $N$ \cite{Yip:2011zz,Cazalilla:2014wfa}; 
	i.e., the number of Nambu--Goldstone modes is equal to that of broken 
	generators and they all enjoy a linear dispersion.  	
	This can be argued as follows. According to 
	\cite{Schafer:2001bq,Watanabe:2012hr,Hidaka:2012ym}, the number of 
	Nambu--Goldstone modes must be equal to the number of broken generators 
	if $\akakko{[Q_a,Q_b]}=0$ for all pairs of broken generators $\{Q_i\}$. 
	In the case of $N$-component fermions with even $N$, the fact that the 
	coset $\SU(N)/\Sp(N)$ is a \emph{symmetric space} \cite{Peskin:1980gc} implies 
	that a commutator of broken generators is a linear combination of \emph{unbroken} 
	generators. Then, if there is a nonzero density of $\Sp(N)$ charges in the ground state, 
	it breaks $\Sp(N)$ and contradicts the assumption of unbroken $\Sp(N)$ symmetry. 
	Hence $\akakko{[Q_a,Q_b]}=0$.   
	\item 
	In two-color QCD, a Wess--Zumino--Witten term proportional to 
	$\epsilon^{\mu\nu\rho\sigma}$ is necessary to account for the axial 
	anomaly at the level of the chiral Lagrangian 
	\cite{Duan:2000dy,Lenaghan:2001sd}.  The same term can emerge 
	in our effective theory as well (in $3+1$ dimensions) at the cost of parity, but 
	this term is fourth order in derivatives and can be safely neglected at low energy. 
	\item 
	Suitable extensions of Eq.~\eqref{eq:Lmain} to the imbalanced case were 
	thoroughly discussed in \cite{Kogut:1999iv,Kogut:2000ek,
	Splittorff:2000mm,Brauner:2006dv,Kanazawa:2009ks} 
	in the context of two-color QCD. 
\end{itemize}
Having introduced EFT, we are in a position to compute low-energy observables. 
We calculate the susceptibility
\ba
	\chi_{AB}(j) & \equiv \lim_{\forall j_A\to 0} \lim_{V\to \infty}
	\frac{1}{\beta V}\frac{\der^2}{\der j_A \der j_B}\log Z 
\ea 
from both the microscopic action and EFT.
In the microscopic theory, we have
\ba
	\chi_{AB}(j) &= \frac{\delta_{AB}}{2} 
	\int_0^\infty \!\!\! \dd \Lambda\,
	\frac{\Lambda^2-j^2}{(\Lambda^2+j^2)^2}R_1(\Lambda)\,. 
	\label{eq:chiR}
\ea
On the EFT side, we find that the leading infrared singularity 
as $j\to 0$ is given by
\ba
	\chi_{AB}(j) & \simeq - \delta_{AB}\,\hat{\chi}\,\log j  \,,
	\label{eq:chiab}
\ea
with 
\ba
	\hat{\chi} & \equiv \frac{1}{128\pi^2} \frac{\Phi^2}{F^4}
	\bigg[
	\frac{(N-4)(N+2)}{8N} \frac{1}{v^3} + 
	\frac{4}{v\tilde{v}(v+\tilde{v})} \frac{F^2}{f^2} 
	\bigg]
	\label{eq:chihat}
\ea
at $D=3$ and $T=0$.
The derivations of Eqs.~\eqref{eq:chiR}, \eqref{eq:chiab}, and \eqref{eq:chihat} are given in \autoref{ap:der}.
The infrared divergence in Eq.~\eqref{eq:chiab} must be accounted for by 
Eq.~\eqref{eq:chiR} as well, i.e., for small $j$ 
\ba
	\int_0^\infty \!\!\! \dd \Lambda\,
	\frac{\Lambda^2-j^2}{(\Lambda^2+j^2)^2}R_1(\Lambda)
	\simeq -2 \hat\chi \log j \,.
\ea
This constrains the possible form of $R_1(\Lambda)$. 
We note that the constant part of 
$R_1(\Lambda)$ does not contribute to the integral  
\cite{Smilga:1993in}, since $\int_0^\infty\dd x\frac{x^2-1}{(x^2+1)^2}=0$. 
A logarithmic divergence could be reproduced if 
$R_1(\Lambda)-R_1(0)$ is linear in $\Lambda$ near the origin.  
Thus we finally obtain 
\ba
	\lim_{j\to 0} R_1(\Lambda) = \frac{2}{\pi}\Phi +2\hat\chi \Lambda + o(\Lambda)  .
	\label{eq:SSrel}
\ea
This is the main result of this section. Equation 
\eqref{eq:SSrel} presents a condensed-matter 
analogue of the Smilga--Stern relation in QCD \cite{Smilga:1993in}.
This relation holds at $T=0$ in $D=3$ dimensions for $N\geq 4$ even. 
Derivation of a similar formula for $N=2$ is left as an interesting open 
problem. Probably this can be handled by means of the supersymmetric method 
along the lines of \cite{Osborn:1998qb,Toublan:1999hi}.  
In $D=2$, the infrared singularity is even stronger and 
$\chi_{AB}(j)$ diverges as $\sim 1/\sqrt{j}$. This implies 
\ba
	R_1(\Lambda) - R_1(0) & \propto j^\alpha \Lambda^{\frac{1}{2}-\alpha}
\ea
up to the scale $\Lambda\sim j$, for an arbitrary $0\leq \alpha \leq1/2$.  
This is all we can say about the form of $R_1$ in $2$ dimensions.

\section{\label{sc:rmt}Random matrix theory}

Although not explicitly shown in Eq.~\eqref{eq:Lmain},  
there are infinitely many terms in the effective Lagrangian  
and it is imperative to organize them in a consistent manner. 
One way to do this is to employ a counting scheme where 
the derivative  ($\der_{\tau}$, $\nabla$) and the mass 
term $m_{\pi,\phi}$ are treated as small quantities of the same order. 
However, there is yet another way of organizing the expansion 
\cite{Gasser:1987ah}. 
Suppose the system is put in a box of linear extent $L$ and 
assume a counting scheme 
\ba
	\begin{split}
		& \qquad  
		\der_\tau \sim \nabla \sim \frac{1}{L}\sim T \sim \calO(\epsilon) \,,
		\\
		& j \sim \calO(\epsilon^{D+1}) \quad \text{and}\quad 
		m_{\pi,\phi} \sim \calO(\epsilon^{\frac{D+1}{2}})\,.
	\end{split}
	\label{eq:epregime}
\ea 
This is called the \emph{$\epsilon$-regime} \cite{Gasser:1987ah}.  
This can be realized by taking the combined limit 
$T\to 0$, $L\to\infty$ and $j\to 0$ 
keeping $\beta V \Phi j \sim 1$.  
In this expansion, the leading term is given by the mass term in 
Eq.~\eqref{eq:Lmain} while all the rest are suppressed by additional 
powers of $\epsilon$, implying that the space-time-dependent part of 
the Nambu--Goldstone modes is suppressed relative to the zero mode 
$\tilde\Sigma=\text{const}$. 
This leads us to an intriguing observation that  
the partition function at leading order of the $\epsilon$-expansion 
reduces to just a finite-dimensional integral over the coset space: 
\ba
	Z & = \int\limits_{\U(N)/\Sp(N)}\!\!\!\!\!\!\!\! \dd \tilde\Sigma\ 
	\exp\big[ \! -\beta V\Phi \,\Re\tr(J \tilde\Sigma)  \big] \,,
	\label{eq:zeroEFT}
\ea
which can be computed analytically \cite{Smilga:1994tb,Nagao:2000qn}. 

A more intuitive way of understanding this dramatic reduction is as 
follows. For $D>1$, the counting \eqref{eq:epregime} implies that 
a separation of scales
\ba
	\beta \ll \frac{1}{m_{\pi,\phi}} 
	\quad \text{and} \quad 
	L\ll \frac{v}{m_\pi},\ \frac{\tilde{v}}{m_\phi}
\ea
holds. This means that the box size is much shorter than the correlation 
lengths in both temporal and spatial directions, so that only zero modes 
of the Nambu--Goldstone modes contribute to the partition function.  
To avoid confusion, we stress that the domain of validity 
for the partition function \eqref{eq:zeroEFT} does \emph{not} overlap with the domain where 
the Banks--Casher-type relation \eqref{eq:newBC} and the 
Smilga--Stern-type relation \eqref{eq:SSrel} hold.  The latter two assume 
that $j\to0$ is taken \emph{after} $\beta V\to \infty$. This is different from 
the $\epsilon$-regime where the two limits must be taken simultaneously. 

Since the form of the partition function \eqref{eq:zeroEFT} is totally fixed 
by global symmetries,  it embodies the \emph{universal} nature of the system. 
Namely, any theory undergoing the same pattern of symmetry breaking 
should reduce to the same partition function in the $\epsilon$-regime, regardless of 
all the complex details of the microscopic Lagrangian.   
This reasoning suggests that the sigma model 
representation \eqref{eq:zeroEFT} may result from a much simpler and 
tractable model. Indeed it has been shown by Verbaarschot 
\emph{et al.}~\cite{Verbaarschot:1993pm,Verbaarschot:1994qf,Verbaarschot:1994ia,Halasz:1995qb} 
in the context of QCD that Eq.~\eqref{eq:zeroEFT} can be reproduced exactly from 
the random matrix theory (RMT)%
\footnote{The connection between RMT and sigma models has been 
discussed in quite general contexts; see e.g., \cite{Verbaarschot:1985jn,Efetov:1997fw}.}  
\ba
	Z_{\rmt} & = \!\! \int\limits_{\RR^{n\times n}} \!\!\!\dd \hat{W}\ 
	{\det}^{N/2} \! \begin{pmatrix}
		\hat{j} & \hat{W} \\ -\hat{W}^T & \hat{j}
	\end{pmatrix}\exp\mkakko{- \frac{n}{2}\tr \hat{W}\hat{W}^T } ,
	\label{eq:chGOE}
\ea
where $\hat{W}$ is a real $n\times n$ matrix and the hat $\hat{}$ is attached 
to dimensionless quantities.  In the $n\to\infty$ limit 
with $n\hat{j}=\calO(1)$, $Z_{\rmt}$ reduces to \eqref{eq:zeroEFT} 
if we identify
\ba
	\beta V \Phi J \Leftrightarrow  n\hat{j} I\,. 
	\label{eq:jmap}
\ea
Equation \eqref{eq:chGOE} is called the \emph{chiral Gaussian orthogonal ensemble}  
(chGOE) which corresponds to Class BDI in the ten-fold symmetry classification of 
RMT \cite{Zirnbauer:1996zz,Altland:1997zz}.%
\footnote{The reader may find the $2\times 2$ block structure of \eqref{eq:chGOE} in the `particle-hole' space to be reminiscent of the Bogoliubov--de Gennes ensemble of random matrices 
\cite{Altland:1996zz,Altland:1997zz}. To avoid confusion, 
let us emphasize that our RMT (chGOE) has 
\emph{no} fluctuating components in the 
particle-particle and hole-hole sector --- namely, 
the Hubbard--Stratonovich transformation leading to \eqref{eq:S'} 
was performed only in the particle-hole channel.} 
While chGOE was originally proposed to describe 
the Dirac operator spectra in two-color QCD, 
it can equally be applied to multi-component Fermi 
gases due to the coincidence of the global symmetry breaking 
pattern, $\U(1)\times\SU(N)\to\Sp(N)$.  
The only notable distinction is that 
$\U(1)$ is violated by quantum anomaly in QCD but not in Fermi gases, 
which is reflected in the form of $\hat{W}$: it is a rectangular matrix in 
applications to QCD but must be a square matrix in our case. 

A notable consequence of the above equivalence between RMT and the 
$\epsilon$-regime EFT is that the statistical correlations of the near-zero 
singular values of $W$ (in the full theory) and $\hat{W}$ 
(in RMT) on the scale of average level spacing should agree exactly.  
This is an example of \emph{spectral universality} that emerges in a variety of physical 
systems \cite{Guhr:1997ve}.  In the model \eqref{eq:chGOE}, 
the average level spacing near zero is of order $\sim 1/n$,  so the universal 
behavior is manifested in the singular value spectrum of $\hat{W}$ 
(denoted as $\{\hat\Lambda_n\}$) on the scale $\sim1/n$. 
This leads us to define the so-called 
\emph{microscopic spectral density} \cite{Verbaarschot:1993pm}
\ba
	\rho_{\scriptscriptstyle\text{RMT}}(\lambda) & \equiv \lim_{n\to\infty} \frac{1}{n}
	\bigg\langle
		\sum_n \delta\bigg(\frac{\lambda}{n} - \hat\Lambda_n \bigg)
	\bigg\rangle. 
\ea
In chGOE, $\rho_{\scriptscriptstyle\text{RMT}}(\lambda)$ has been computed 
analytically at $\hat{j}=0$ in \cite{Verbaarschot:1994ia} and for general $\hat{j}\ne0$ 
in \cite{Nagao:2000cb}.  
Now, based on the correspondence between RMT and EFT [cf.\,\eqref{eq:jmap}], 
we expect that $\rho(\lambda)$ defined in the full theory as 
\ba
	\rho(\lambda) & \equiv \lim_{\beta V\to\infty}
	\frac{1}{\beta V \Phi}\bigg\langle
		\sum_n \delta\bigg(\frac{\lambda}{\beta V \Phi} - \Lambda_n \bigg)
	\bigg\rangle
	\label{eq:msd}
\ea
must coincide with $\rho_{\scriptscriptstyle\text{RMT}}(\lambda)$ exactly.%
\footnote{Note the difference from the definition \eqref{eq:R1} of $R_1(\Lambda)$. 
The microscopic spectral density $\rho(\lambda)$ 
looks at a much finer structure of the spectrum than the spectral density $R_1(\Lambda)$ does.}
This coincidence should also occur for higher-order correlation functions and 
the smallest singular-value distribution $P(\lambda_{\rm min})$. 
The latter was analytically computed for chGOE by various authors 
\cite{Edelman1988,Edelman1991,Forrester1993,
Damgaard:2000ah,Akemann:2014cna,Wirtz:2015oma}.  
In the case of QCD, a quantitative agreement between the Dirac spectrum 
in QCD and the prediction of RMT for $\rho(\lambda)$ and $P(\lambda_{\rm min})$ 
has been firmly established through Monte Carlo simulations \cite{BerbenniBitsch:1997tx} 
(see \cite{Verbaarschot:2000dy,Verbaarschot:2009jz} for reviews). 
Before proceeding, let us give a couple of comments regarding $\rho(\lambda)$: 
\begin{itemize}
	\item 
	One can define the microscopic spectral density only in the symmetry-broken 
	phase. In the symmetric phase, there is no small singular values of order 
	$1/V$ and the correspondence to RMT is lost. 
	\item 
	In numerical simulations in the $\epsilon$-regime, 
	one needs to rescale the spectrum 
	of dimensionless singular values so as to match 
	$\rho_{\scriptscriptstyle\text{RMT}}(\lambda)$.  This procedure 
	allows us to extract the value of $\Phi$ accurately. On the other hand, 
	the Banks--Casher-type relation $\lim_{j\to 0} R_1(0)=\frac{2}{\pi}\Phi$ also 
	gives $\Phi$.  The values of $\Phi$ obtained in these ways should 
	agree with each other, since $\Phi$ is a physical observable that enters 
	the low-energy effective theory \eqref{eq:Lmain}. Note however that 
	these measurements cannot be done simultaneously, as they 
	have non-overlapping domains of validity.  In practical simulations, 
	the volume is necessarily finite and any measurement is afflicted with 
	finite-volume effects. Of the two methods, one should use the one 
	that receives smaller finite-volume corrections in a given setting.    
	\item 
	Once the symmetry of the action is modified by external 
	perturbations, the corresponding RMT can change 
	from chGOE to something else.  For instance, 
	coupling of fermions to an external gauge field would make 
	the matrix $W$ complex. The appropriate RMT 
	is now the chiral Gaussian unitary ensemble (chGUE)  
	\cite{Shuryak:1992pi,Verbaarschot:1994qf}, which has 
	complex matrix elements.  
	In principle one can investigate a crossover between 
	chGOE and chGUE in numerical simulations. 
	\item 
	Yet another perturbation of physical importance is a 
	species imbalance (or \emph{polarization}). Let us take $N=2$ 
	for illustration. If the chemical potential for up($\uparrow$) fermions 
	is detuned from that of down($\downarrow$) fermions by an 
	amount $\delta\mu\ne 0$, the partition function 
	can no longer be expressed in terms of 
	a single operator $WW^\dagger$ as in \eqref{eq:Zf}.   
	Instead, one has to handle a complex eigenvalue 
	spectrum of a non-Hermitian operator $W_\uparrow W_\downarrow$ with 	
	$W_\uparrow\ne W_\downarrow^\dagger$.%
	\footnote{The same situation arises when the masses of $N$ components are unequal.}     
	We are then forced to adopt a 
	non-Hermitian extension of chGOE (or a chiral extension of 
	the so-called real Ginibre ensemble \cite{Ginibre:1965zz}) 
	to describe universal correlations of \emph{complex} eigenvalues 
	of $W_\uparrow W_\downarrow$.%
	\footnote{Note that, in the $\epsilon$-regime, 
	$\delta\mu\sim \calO(\epsilon^{\frac{D+1}{2}})$ 
	must go to zero in the thermodynamic limit 
	\cite{Akemann:2004dr}. This means that RMT cannot be 
	used to describe phase transitions that occur 
	in the $V\to\infty$ limit with $\delta\mu\ne0$ fixed.
	} 
	Such an extension of chGOE has already been 
	thoroughly studied and even analytically solved  
	in \cite{Akemann:2009fc,Kanazawa:2009en,Akemann:2010tv}, 
	aiming at applications to two-color QCD with baryon chemical 
	potential. Based on the universality of RMT, 
	we believe that the level statistics of the non-Hermitian chGOE 
	should apply to the imbalanced Fermi gases as well.   
\end{itemize}

\section{Numerical simulation}\label{sc:NS}

We checked a few of the theoretical findings in the former sections 
by the path-integral Monte Carlo simulation, which is familiar in lattice QCD \cite{Rothe:1992nt}.
The Monte Carlo configurations are generated on the basis of the measure \eqref{eq:Zf}, and then the statistical average over configurations is taken.
The operator \eqref{eq:W} is discretized on a (3+1)-dimensional lattice as
\ba
	\begin{split}
	 W_{x,x'} & = \frac{1}{a}\big[
	 \delta_{x,x'} - \{1+ga\phi(x)\} \ee^{\mu a} \delta_{x-e_\tau,\,x'}
	 \big]
	 \\
	 & \quad - \frac{1}{2ma^2} \sum_{i=x,y,z}
	 (\delta_{x+e_i,\,x'} + \delta_{x-e_i,\,x'} - 2\delta_{x,x'}) \, ,
	\end{split}
\ea
where $e_i$ is the unit lattice vector in the $i$-direction and $a$ is the lattice constant \cite{Chen:2003vy}.
Boundary conditions are periodic in spatial directions and antiperiodic in the $\tau$-direction.
The particle mass and the chemical potential are fixed at $2ma = 10$ and $\mu = 0$, respectively.

\begin{figure}[t]
	\includegraphics[width=8.5cm]{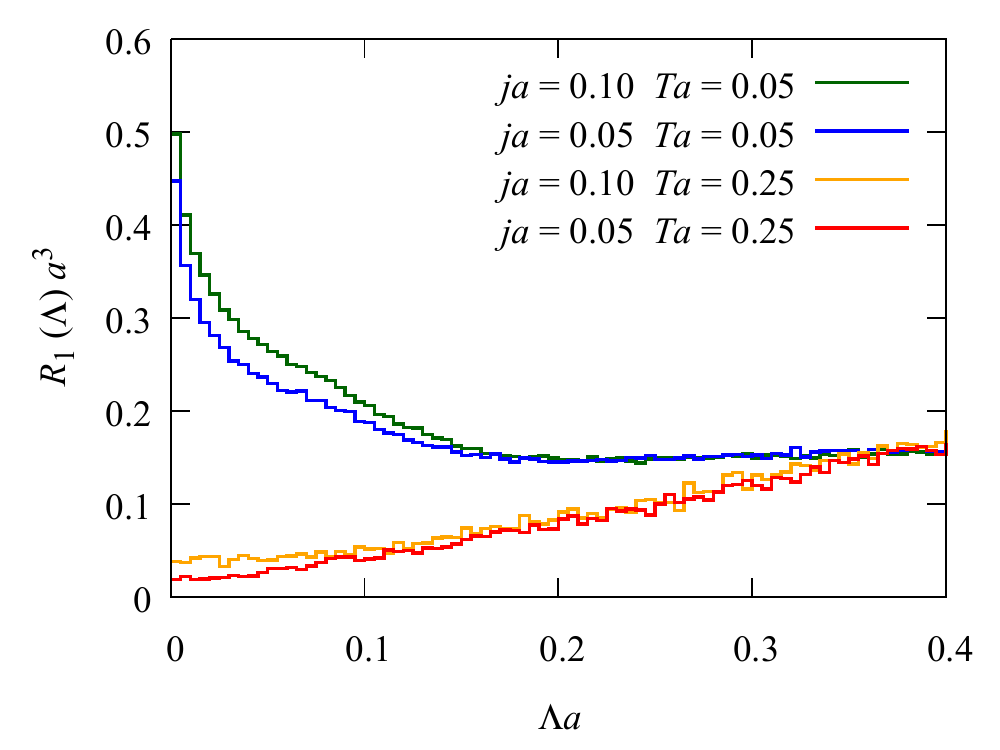}
		\caption{\label{fg_R}
		Spectral density $R_1(\Lambda)$ in the lattice simulation for $N=2$ and $g/a^2=1$. 
		The averages over 500 configurations are shown.
	}
\end{figure}

We numerically computed the singular values $\Lambda_n$, i.e., the square roots of the eigenvalues of the matrix $WW^\dagger$.
The configurations for $N=2$ were generated by the Hybrid Monte Carlo algorithm \cite{Duane:1987de}.
To measure the spectral density \eqref{eq:R1}, 
we performed simulations at $L/a=4$, 6, and 8, and then extrapolated the results to the infinite volume limit.
The obtained spectral density is shown in Fig.~\ref{fg_R}.
At a low temperature ($Ta=0.05$), the spectral density has a peak at $\Lambda=0$ and $R_1(0)$ is clearly nonzero.
From the Banks-Casher-type relation \eqref{eq:newBC}, this indicates a nonzero fermion condensate in a superfluid phase.
At a high temperature ($Ta=0.25$), the spectral density is a slowly increasing function and $R_1(0)$ is close to zero.
This indicates a normal phase.

\begin{figure}[t]
	\includegraphics[width=8.5cm]{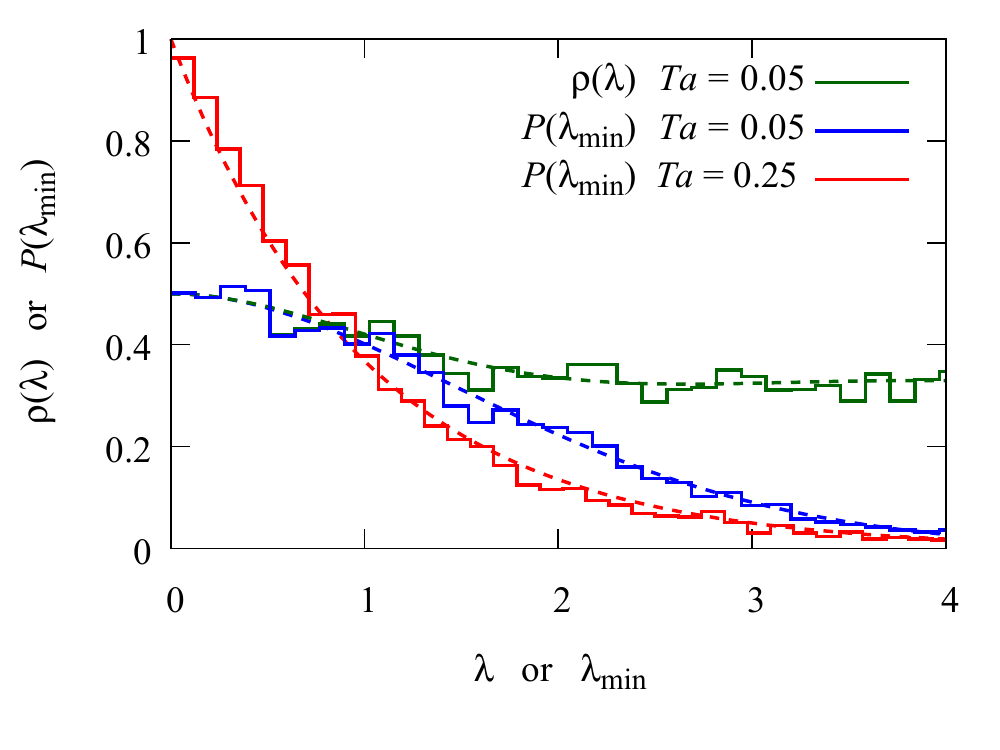}
		\caption{\label{fg_P}
		Microscopic spectral density $\rho(\lambda)$ and the smallest singular-value distribution 
		$P(\lambda_{\rm min})$ in the quenched lattice simulation with $g/a^2=3$.
                The green, blue, and red broken curves are the predictions \eqref{eq:rhoquench} and \eqref{eq:Pquench} by RMT 
                and the Poisson statistics \eqref{eq:PPoisson}, respectively.
                The averages over 5000 configurations are shown.
	}
\end{figure}

We also checked the correspondence to RMT in a small finite volume $V/a^3 = 4^3$.
To increase the number of configurations, we adopted the quenched approximation, which is frequently used in lattice QCD to reduce the computational cost \cite{Rothe:1992nt}.
In the quenched approximation, the fermion determinant in the measure \eqref{eq:Zf} is neglected.
The measure is thus given by a product of independent Gaussian weights for $\phi(x)$, 
which helps to speed up the simulation extremely.
Since quenched configurations are independent of $j$, singular-value distributions have no dependence on $j$.
In Fig.~\ref{fg_P}, the microscopic spectral density $\rho(\lambda)$ and the smallest singular-value distribution $P(\lambda_{\rm min})$ are shown.
In the quenched chGOE with trivial topology ($\nu=0$), they are analytically 
given by \cite{Verbaarschot:1994ia,Forrester1999}
\ba
	\begin{split}
		\rho_{\scriptscriptstyle\text{RMT}}(\lambda) = & 
		\frac{\lambda}{2} \{ J_0(\lambda)^2 + J_1(\lambda)^2 \} 
		\\
		& + \frac{1}{2} J_0(\lambda) 
		\bigg( 1 - \int_0^\lambda \!\!\! \dd x ~J_0(x) \bigg)
	\end{split}
	\label{eq:rhoquench}
\ea
and \cite{Forrester1993}
\ba
	P_{\scriptscriptstyle\text{RMT}}(\lambda_{\rm min}) & = 
	\frac{2+\lambda_{\rm min}}{4} 
	\exp \left( -\frac{\lambda_{\rm min}^2}{8} 
	- \frac{\lambda_{\rm min}}{2} \right) ,
	\label{eq:Pquench}
\ea
respectively, which are drawn in Fig.~\ref{fg_P} for comparison.
Although these analytical solutions of RMT are parameter free if $\Phi$ is known, here $\Phi$ is treated as a fitting parameter.
At a low temperature ($Ta=0.05$), the lattice simulation nicely reproduces 
the predictions of RMT.
At a high temperature ($Ta=0.25$), the lattice simulation deviates from RMT 
and approaches the Poisson distribution 
\ba
	P(\lambda_{\rm min}) = \exp(-\lambda_{\rm min})\,,
	\label{eq:PPoisson}
\ea
which signals absence of a level correlation.

\section{Summary and perspective}\label{sc:SP}

In this work, we studied multi-component fermionic superfluids and derived a number of rigorous results 
by using theoretical methods that hardly appear in conventional studies of 
nonrelativistic systems but are well established in the field of Quantum Chromodynamics (QCD). 
By relating the order parameters of spontaneous symmetry breaking to the singular-value spectrum 
of a single operator $W$ [Eq.~\eqref{eq:W}] we derived a nonrelativistic analog of the Banks--Casher relation 
in QCD, which enables us to extract the bifermion condensate $\akakko{\psi\psi}$ from the spectrum 
reliably. Furthermore we have shown, through a spectral analysis of $W$, 
that $\U(1)$ and $\SU(N)$ symmetry of the $N$-component Fermi gas must be restored/broken 
simultaneously.  This imposes a strong constraint on the phase diagram by precluding 
intermediate phases where either $\U(1)$ or $\SU(N)$ is broken and the other is not. 

We also developed a low-energy effective theory of Nambu--Goldstone modes 
in the superfluid phase for general even $N$, and rigorously derived a formula which 
expresses the slope of the spectral density near zero in terms of low-energy 
constants that enter the effective Lagrangian.  This is an analog of the 
Smilga--Stern relation in QCD.  In addition, we pointed out that the effective theory 
can be mapped to a random matrix theory in the $\epsilon$-regime.  
From this correspondence we found an analytical formula for the spectral density 
near zero. This provides us with a novel, numerically clean method to 
extract the bifermion condensate through fitting to the numerical data of the spectrum.  
We confirmed these analytical calculations by the path-integral Monte 
Carlo simulations. 

It should be emphasized that the analysis in this paper involves no uncontrolled 
approximations and is valid under quite general conditions for temperature, density and 
the interaction strength, as long as the path-integral measure is positive definite.  
Our results can be used to benchmark other theoretical methods. 

There are various future directions. 
The multi-component Hubbard model can be studied in the same manner.
This may add to our knowledge of the rigorous results of the Hubbard model. 
It is also worthwhile to extend the present work to other cases of $N$; in particular,
the symmetry analysis based on the multi-point correlation functions in Sec.~\ref{sc:u1} to $N\ge 6$,
the Smilga--Stern-like relation in Sec.~\ref{sc:SS} to $N=2$,
and the numerical checks of correspondence to RMT in Sec.~\ref{sc:NS} to unquenched $N\ge 2$.
The extension of our framework to theories with odd $N$ or repulsive 
interactions is a more challenging problem.

\begin{acknowledgments}
TK was supported by the RIKEN iTHES project. 
AY was supported by JSPS KAKENHI Grants Number 15K17624. 
TK is grateful to Shinsuke M. Nishigaki and Shun Uchino for helpful discussions.  
The numerical simulations were carried out on SX-ACE in Osaka University.
\end{acknowledgments}

\appendix

\section{\label{ap:free}Spectral density in a free theory} 

In the free limit $g\to 0$, the spectral density can be obtained analytically. 
The spectral density is independent of $j$ in a free theory. 
At $T=0$,
\ba
	& \quad R_1(\Lambda) 
	\notag
	\\
	& = \frac{1}{\beta V} 
	\bigg\langle \sum_{n}\delta(\Lambda_n - \Lambda) \bigg\rangle  
	\notag
	\\
	& = \frac{2 \Lambda}{\beta V}  
	\tr\kkakko{\delta(WW^\dagger - \Lambda^2)} 
	\notag
	\\
	& = \frac{2 \Lambda}{\beta V}  
	\tr\kkakko{\delta\bigg( 
		-\der_\tau^2 + \mkakko{-\frac{\nabla^2}{2m}-\mu}^2 - \Lambda^2
	\bigg)}
	\notag 
	\\
	& = 2 (2m)^{D/2} \Lambda \! \int\!\frac{\dd \omega}{2\pi}\! 
	\int\!\! \frac{\dd^D q}{(2\pi)^D}
	~\delta\mkakko{
		\omega^2 + (q^2-\mu)^2 - \Lambda^2 
	}
	\notag
	\\
	& = \frac{C_D}{(2\pi)^D} (2m)^{D/2} \Lambda~ \times 
	\notag
	\\
	& \qquad \times 
	\int\!\frac{\dd \omega}{2\pi}\!\int_0^\infty \!\!\! \dd r~r^{(D-2)/2}
	~\delta\mkakko{
		\omega^2 + (r - \mu)^2 - \Lambda^2 
	} 
	\notag 
	\\
	& = \frac{C_D}{(2\pi)^{D+1}} (2m)^{D/2} \Lambda ~ \times
	\notag
	\\
	& \qquad \times 
	\int_0^\infty \!\!\! \dd r~r^{(D-2)/2}
	~\frac{
		\Theta\big( \Lambda^2 - \mkakko{r - \mu}^2 \big)
	}{
		\sqrt{ \Lambda^2 - \mkakko{r - \mu}^2 } 
	} 	\,,
	\label{eq:Rlas}
\ea
where $\Theta(x)$ is a step function and $C_1=2$, $C_2=2\pi$ and $C_3=4\pi$. 
The integrand of \eqref{eq:Rlas} is nonzero for  
$(r-\mu)^2\leq \Lambda^2$, i.e., $\mu-\Lambda \leq r \leq \mu + \Lambda$. 
We divide the $(\mu,\Lambda)$-plane 
into two regions: $\mu\geq\Lambda$ and $\Lambda\geq \mu$.    
\vspace{5pt}
\\
$\bullet$ \underline{Case I:~$\mu\geq\Lambda$}
\vspace{3pt}
\\
Writing $r=\mu+\Lambda \cos\theta$ ($0\leq\theta\leq \pi$), we obtain
\ba
	& \quad R_1(\Lambda) 
	\notag
	\\
	& = \frac{C_D}{(2\pi)^{D+1}} (2m)^{D/2} \Lambda \!
	\int_{\mu-\Lambda}^{\mu+\Lambda} \!\!\! \dd r \frac{r^{(D-2)/2}}
	{
		\sqrt{ \Lambda^2 - \mkakko{r - \mu}^2 } 
	}
	\notag 
	\\
	& = \frac{C_D}{(2\pi)^{D+1}} (2m)^{D/2} \Lambda \! 
	\int_0^\pi \!\! \dd\theta \mkakko{\mu+\Lambda \cos\theta}^{(D-2)/2}
	\notag 
	\\
	& = \left\{\begin{array}{ll}
		\displaystyle \frac{m}{2\pi}\Lambda & [D=2]
		\vspace{3pt}
		\\
		\displaystyle 
		\frac{(2m)^{3/2}}{2\pi^3}\Lambda\sqrt{\mu+\Lambda} 
		\EE\mkakko{\frac{2\Lambda}{\mu+\Lambda}} & [D=3]
	\end{array}
	\right.  ,
	\label{eq:rhoI}
\ea
where $\EE(x)\equiv\int_0^{\pi/2}\dd\theta\sqrt{1-x \sin^2\theta}$ 
is the complete elliptic integral of the second kind. 
Interestingly, $R_1(\Lambda)$ has no $\mu$-dependence 
for $D=2$. 
\vspace{5pt}
\\
$\bullet$ \underline{Case II:~$\Lambda\geq \mu$} 
\vspace{3pt}
\ba
	& \quad R_1(\Lambda) 
	\notag
	\\
	& = \frac{C_D}{(2\pi)^{D+1}} (2m)^{D/2} \Lambda \!  
	\int_{0}^{\mu+\Lambda} \!\!\! \dd r \frac{r^{(D-2)/2}}
	{
		\sqrt{ \Lambda^2 - \mkakko{r - \mu}^2 } 
	}
	\notag
	\\
	& = \frac{C_D}{(2\pi)^{D+1}} (2m)^{D/2} \Lambda 
	~\times
	\notag
	\\
	& \qquad \times 
	\int_0^{\Xi} \!\!\dd\theta 
	\mkakko{\mu+\Lambda \cos\theta}^{(D-2)/2}
	\Big|_{\Xi:=\cos^{-1}(-\frac{\mu}{\Lambda})}
	\notag
	\\
	& = \left\{\begin{array}{ll}
		\displaystyle \frac{m}{2\pi^2}\Xi \Lambda & [D=2]
		\vspace{3pt}
		\\
		\displaystyle 
		\frac{(2m)^{3/2}}{2\pi^3}\Lambda\sqrt{\mu+\Lambda} 
		\EE\mkakko{\frac{\Xi}{2}\bigg|\frac{2\Lambda}{\mu+\Lambda}} & [D=3]
	\end{array}
	\right. ,
	\label{eq:rhoII}
\ea
	where $\EE(\varphi|x)\equiv\int_0^{\varphi}\dd\theta\sqrt{1-x \sin^2\theta}$ 
	is the incomplete elliptic integral of the second kind. 
	If we formally set $\Xi=\pi$,  
	\eqref{eq:rhoII} reduces to \eqref{eq:rhoI}. 

\begin{figure}[t]
	\includegraphics[width=8.5cm]{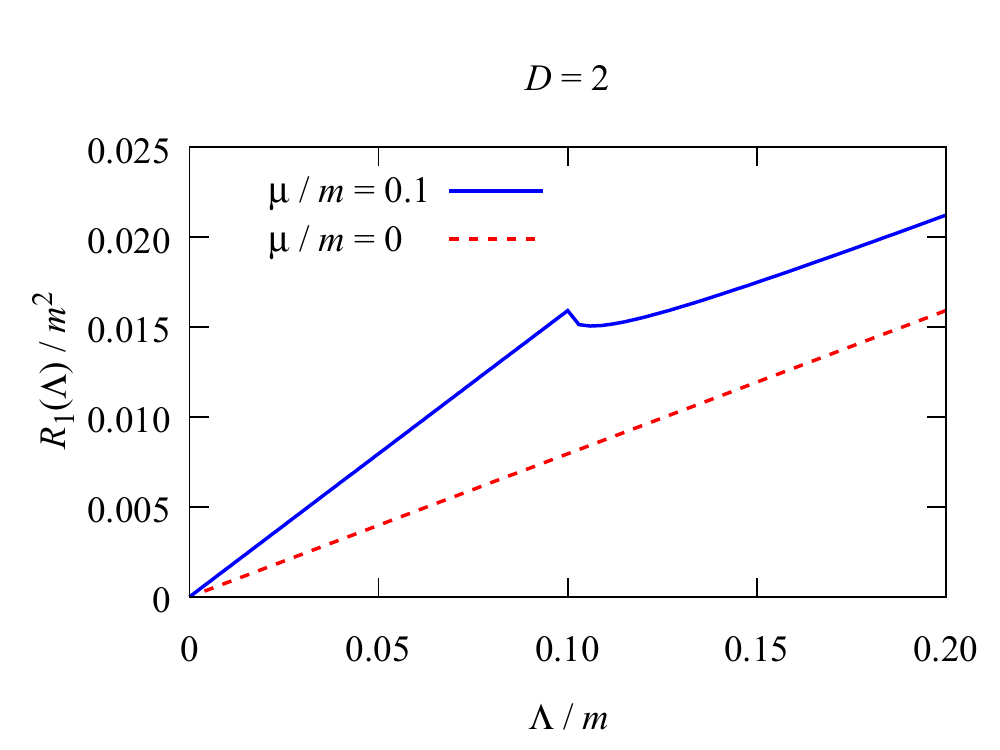}
	\vspace{7pt}
	\\
	\includegraphics[width=8.5cm]{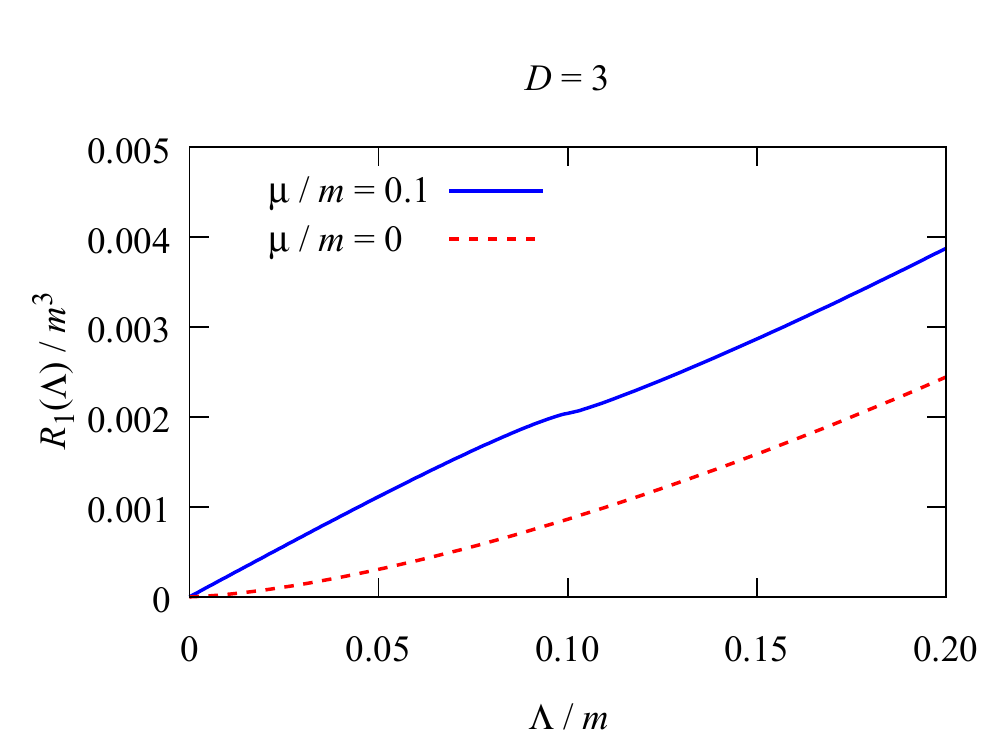}
		\caption{\label{fg_freerho}
		Spectral density in the non-interacting limit at $T=0$ 
		in $2$ (\textbf{top}) and $3$ (\textbf{bottom}) 
		spatial dimensions. 
	}
\end{figure}

Figure \ref{fg_freerho} displays $R_1(\Lambda)$. 
In the limit $\mu\to 0$ (or $\Lambda \gg \mu$), 
one finds from \eqref{eq:rhoII} 
\ba
	R_1(\Lambda) \sim 
	\left\{\begin{array}{ll}
		\displaystyle \frac{m}{4\pi}\Lambda & ~[D=2]
		\vspace{3pt}
		\\
		\displaystyle 
		c \frac{(2m)^{3/2}}{2\pi^3}\Lambda\sqrt{\Lambda} 
		& ~[D=3]
	\end{array}
	\right.  ,\ 
	\label{eq:asym1}
\ea
where $\displaystyle c\equiv \EE\mkakko{\frac{\pi}{4}\bigg|2}=0.59907\dots$ 
is a numerical constant. In contrast, for $\Lambda \ll \mu$,
\eqref{eq:rhoI} becomes 
\ba
	R_1(\Lambda) \sim
	\left\{\begin{array}{ll}
		\displaystyle \frac{m}{2\pi}\Lambda & ~[D=2]
		\vspace{3pt}
		\\
		\displaystyle \frac{\pi}{2}
		\frac{(2m)^{3/2}}{2\pi^3}\sqrt{\mu}\,\Lambda  
		& ~ [D=3]
	\end{array}
	\right.   . 
	\label{eq:asym2}
\ea
At $T>0$, the continuous frequency $\omega$ should be replaced with 
the Matsubara frequency $\omega_n=(2n+1)\pi T$. This means, 
in a plane basis, 
\ba
	WW^\dagger = \omega_n^2 + \mkakko{\frac{\bm{p}^2}{2m}-\mu}^2 
	\geq \omega_0^2=(\pi T)^2 \,. 
\ea
Therefore $R_1(\Lambda)$ vanishes identically on the interval 
$\Lambda\in [0,\pi T]$.

\section{\label{ap:cor}Correlation functions}
In this appendix, we summarize technical formulas of the correlation functions used in Sec.~\ref{sc:u1}.
The propagators for the theory \eqref{eq:Zf} read 
\begin{subequations}
		\ba
			\contraction{}{\psi}{(x)}{\psi} 
			\psi(x)\psi^\dagger(y) 
			& = \langle x | \frac{\1_N}{W^\dagger W+j^2}
			W^\dagger |y\rangle 
			\\
			\contraction{}{\psi}{(x)}{\psi}
			\psi(x)\psi^T(y) 
			& = \langle x| \frac{jI}{W^\dagger W+j^2} |y\rangle 
			\\
			\contraction{}{\psi}{^*(x)}{\psi}
			\psi^*(x)\psi^T(y) 
			& = \langle x| -\frac{\1_N}{WW^\dagger+j^2}W |y\rangle 
			\\
			\contraction{}{\psi}{^*(x)}{\psi}
			\psi^*(x)\psi^\dagger(y) 
			& = \langle x| - \frac{jI}{WW^\dagger+j^2} |y\rangle .
		\ea
		\label{eq:propa}%
\end{subequations}  
It is a straightforward exercise to evaluate the integrated connected correlator \eqref{eq:cx} for the multiplet $(\Pi^0,\Delta^0,\Pi^A,\Delta^A)$ defined in Eq.~\eqref{eq:multiplet}.
Noting that the disconnected piece is nonzero only for $\Delta^0$, we find%
\begin{subequations}
	\ba
		\C_{\Pi^0} & = 2 \bigg\langle \!\tr \frac{1}{W^\dagger W+j^2} \bigg\rangle ,
		\\
		\C_{\Delta^0} & = 
		2 \bigg\langle \!\tr \frac{W^\dagger W-j^2}{(W^\dagger W+j^2)^2} \bigg\rangle
		+ 2N \bigg\langle \!\bigg(\!\tr \frac{j}{W^\dagger W+j^2}\bigg)^2 \bigg\rangle
		\notag
		\\
		& \quad - 2N \bigg\langle \!\tr \frac{j}{W^\dagger W+j^2} \bigg\rangle^2 ,
		\\
		\C_{\Pi^A} & = 4 \tr(T^A T^A) 
		\bigg\langle \!\tr \frac{1}{W^\dagger W+j^2} \bigg\rangle ,
		\\
		\C_{\Delta^A} & = 4 \tr(T^A T^A) 
		\bigg\langle \!\tr \frac{W^\dagger W-j^2}{(W^\dagger W+j^2)^2} \bigg\rangle. 
	\ea%
\end{subequations}
In deriving these results we used the identity  
\ba
	\tr \frac{j}{WW^\dagger +j^2}=\tr\frac{j}{W^\dagger W+j^2}\,.
\ea
Note that an analogue of this relation for the Dirac operator 
does not hold in QCD because of the index theorem.
The summation over $A$ can be easily taken with the help of 
the identity
\ba
		\sum_A T^A T^A & = \frac{N^2-N-2}{4N}\1_N \,.
		\label{eq:TAsum}
\ea

\section{\label{ap:der}\boldmath Derivation of $\chi_{AB}(j)$}

Here, we derive Eqs.~\eqref{eq:chiR}, \eqref{eq:chiab}, and \eqref{eq:chihat}.
In the microscopic theory, the partition function \eqref{eq:Zf} is modified by the generalized source term \eqref{eq:SJgeneral} as $Z(j) \to Z(J)$.
The susceptibility is
\ba
\begin{split}
	\chi_{AB}(j) & \equiv \lim_{\forall j_A\to 0} \lim_{V\to \infty}
	\frac{1}{\beta V}\frac{\der^2}{\der j_A \der j_B}\log Z(J)
        \\
        & = \lim_{\forall j_A\to 0} \lim_{V\to \infty} \frac{1}{4 \beta V} \int_x\int_y \akakko{\Delta^A(x) \Delta^B(y)}_J
        \\
        & = \frac{\delta_{AB}}{2} 
	\int_0^\infty \!\!\! \dd \Lambda\,
	\frac{\Lambda^2-j^2}{(\Lambda^2+j^2)^2}R_1(\Lambda)\,.  
\end{split}
\ea
This gives Eq.~\eqref{eq:chiR}.
While the average $\akakko{\cdots}_J$ is taken over the measure with a generalized source term \eqref{eq:SJgeneral}, $R_1(\Lambda)$ is given by the original measure \eqref{eq:Zf} in the $j_A\to 0$ limit.

In EFT, we assume $T=0$ for the sake of simplicity.
The $j_A$-dependent part of the source term is 
\ba
	\Re\tr(J \tilde\Sigma) & = j_A \mathcal{V}^A+ \dots  
\ea
with
\ba
	\mathcal{V}^A & \equiv \frac{1}{4F^2}\tr\!\mkakko{T^A\{ T^P,T^Q \}} \pi^P\pi^Q
	+ \frac{1}{2Ff}\phi \pi^A
\ea
in the leading order of $\phi$ and $\pi^A$.
Hence 
\ba
	\chi_{AB}(j) = \frac{\Phi^2}{\beta V}\int_x \int_y 
	\akakko{ \mathcal{V}^A(x)\mathcal{V}^B(y) }_{\text{1-loop}}
\ea
where $\akakko{\mathcal{V}^A}=0$ was used. The subscript 
implies we will perform a one-loop analysis, which is sufficient to see 
the leading infrared behavior. As the cross term 
$\akakko{\pi^P\pi^Q\phi \pi^A}=0$, we get
\ba
	\begin{split}
		& \chi_{AB}(j) 
		\\
		& = \frac{\Phi^2}{8F^4}
		\tr\!\mkakko{T^A\{ T^P,T^Q \}}
		\tr\!\mkakko{T^B\{ T^P,T^Q \}} \times 
		\\
		& \qquad \quad \times 
		\int_p \frac{1}{(\omega^2+v^2\bm{p}^2+m_\pi^2)^2}
		\\
		& \quad + \delta_{AB}\frac{\Phi^2}{4F^2f^2}\int_p 
		\frac{1}{\omega^2+v^2\bm{p}^2+m_{\pi}^2}
		\frac{1}{\omega^2+\tilde{v}^2\bm{p}^2+m_\phi^2} \,,
	\end{split}
	\label{eq:chie}
	\hspace{-20pt}
\ea
with $\int_p\equiv \int \frac{\dd\omega}{2\pi}\int\frac{\dd^D p}{(2\pi)^D}$\,.  
We consult \cite{Toublan:1999hi} to obtain 
\ba
	& \tr\!\mkakko{T^A\{ T^P,T^Q \}}	\tr\!\mkakko{T^B\{ T^P,T^Q \}} 
	\notag
	\\
	& \qquad = \frac{(N-4)(N+2)}{8N}\delta_{AB}\,.  
	\label{eq:Liefactor}
\ea
The momentum integrals in Eq.~\eqref{eq:chie} are divergent in the limit $j\to 0$ for $D\leq 3$. 
At $D=3$, the leading singularity in $j \to 0$ is
\ba
	& \int_p \frac{1}{(\omega^2+v^2\bm{p}^2+m_\pi^2)^2} 
	\simeq - \frac{1}{16\pi^2}\frac1{v^3} \log j   \,,
	\\
	& \int_p \frac{1}{\omega^2+v^2\bm{p}^2+m_{\pi}^2}
	\frac{1}{\omega^2+\tilde{v}^2\bm{p}^2+m_\phi^2}
	\notag
	\\
	& \qquad 
	\simeq - \frac{1}{8\pi^2} \frac{1}{v\tilde{v}(v+\tilde{v})} \log j \,. 
\ea
We used $m_\pi^2=j\Phi/2F^2$ and $m_\phi^2=jN\Phi/f^2$.
Collecting everything, we obtain Eqs.~\eqref{eq:chiab} and \eqref{eq:chihat}.
In the same way one can show $\chi_{AB}(j) \sim 1/\sqrt{j}$ at $D=2$.

\bibliography{paper.bbl}
\end{document}